\documentclass[10pt,twocolumn]{article}

\usepackage{amsmath}
\usepackage{amssymb}
\usepackage{array}
\usepackage{graphicx}

\usepackage{cite}

\usepackage{color}

\usepackage{setspace}
\usepackage[labelfont=bf,labelsep=period,justification=raggedright]{caption}

\makeatletter
\renewcommand{\@biblabel}[1]{\quad#1.}
\makeatother

\date{}

\begin{document}

\begin{flushleft}
{\Large
\textbf{A Cellular Automaton Model for Tumor Dormancy: Emergence of a Proliferative Switch}
}
\\
Duyu Chen$^{1,2}$, Yang Jiao$^{2,3}$, Salvatore
Torquato$^{1,2,4,5,6\ast}$
\\
\bf{1} Department of Chemistry, Princeton University, Princeton, New Jersey 08544, USA
\\
\bf{2} Physical Science in Oncology Center, Princeton University, Princeton, New Jersey 08544, USA
\\
\bf{3} Materials Science and Engineering, Arizona State
University, Tempe, Arizona 85287, USA
\\
\bf{4} Department of Physics, Princeton University, Princeton, New
Jersey 08544, USA
\\
\bf{5} Program in Applied and Computational Mathematics, Princeton
University, Princeton, New Jersey 08544, USA
\\
\bf{6} Princeton Institute for the Science and Technology of
Materials, Princeton University, Princeton, New Jersey 08544, USA
\\
$\ast$ E-mail: torquato@electron.princeton.edu
\end{flushleft}

\section*{Abstract}

Malignant cancers that lead to fatal outcomes for patients may remain dormant for very long periods of time. Although individual mechanisms such as cellular dormancy, angiogenic dormancy and immunosurveillance have been proposed, a comprehensive understanding of cancer dormancy and the ``switch'' from a dormant to a proliferative state still needs to be strengthened from both a basic and clinical point of view. Computational modeling enables one to explore a variety of scenarios for possible but realistic microscopic dormancy mechanisms and their predicted outcomes. The aim of this paper is to devise such a predictive computational model of dormancy with an emergent ``switch'' behavior. Specifically, we generalize a previous cellular automaton (CA) model for proliferative growth of solid tumor that now incorporates a variety of cell-level tumor-host interactions and different mechanisms for tumor dormancy, for example the effects of the immune system. Our new CA rules induce a natural ``competition'' between the tumor and tumor suppression factors in the microenvironment. This competition either results in a ``stalemate'' for a period of time in which the tumor either eventually wins (spontaneously emerges) or is eradicated; or it leads to a situation in which the tumor is eradicated before such a ``stalemate'' could ever develop. We also predict that if the number of actively dividing cells within the proliferative rim of the tumor reaches a critical, yet low level, the dormant tumor has a high probability to resume rapid growth. Our findings may shed light on the fundamental understanding of cancer dormancy.

\section*{Introduction}
Cancer dormancy, the phenomena that the tumor's volume or the number of tumor cells stays at a very low level for a certain period of time before the tumor begins to grow rapidly, has been an outstanding issue in cancer research for many years \cite{Almog2010139,aguirre2007models}. Currently, the mechanisms responsible for the ``switch'' from a dormant state to a rapid growth state for different tumors are not well understood, although it is well known that such a ``switch'' in secondary metastatic tumors can be triggered by the removal of the primary tumor. This could eventually lead to failure of tumor treatment and fatal outcomes for the patient. Therefore, a comprehensive understanding of the ``switch'' from a dormant to a proliferative state is crucial to our fundamental understanding of cancer progression and recurrence and might lead to the development of novel treatments for cancer.

Dormancy has been observed in many types of cancer. This includes tumor dormancy before any metastases take place and the latency of cancer recurrence after therapy. In some cases of pancreatic cancer, the tumor can remain in a benign dormant state for about 20 years \cite{yachida2010distant}. During this time, it is undetectable by conventional clinical methods, and it is only afterwards that the tumor becomes highly malignant and grows aggressively with highly fatal outcomes after about a year. In the cases of breast and prostate cancer, it is reported that $20\%$-$45\%$ of patients will relapse years or decades later after the resection of the primary tumor \cite{karrison1999dormancy,pfitzenmaier2006telomerase,
weckermann2001disseminated}. In addition, recurrence has been observed in brain tumors, which indicates the existence of a large number of micrometastases that are dormant in the presence of the primary tumor \cite{ghiso1999tumor,naumov2006model}.

\begin{figure}[!ht]
\begin{center}
$\begin{array}{c}
\includegraphics[width=0.45\textwidth]{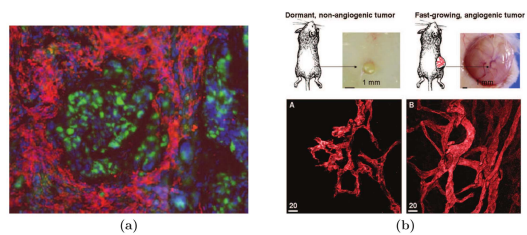}
\end{array}$
\end{center}
\caption{(Color online) (a) Fluorescence micrograph of a breast tumor stained to visualize carcinoma cells (phospho-p53, green) surrounded by macrophages (CD11b, red). Nuclei appear blue (DAPI). Image courtesy of Michael Graham Espey, PhD, National Cancer Institute, NIH (private communication). (b) Representative pictures of dormant and fast-growing tumors and their vascular structure. Reprinted from Cancer Letters, 294, Almog N, Molecular mechanisms underlying tumor dormancy, 139--146, Copyright (2010), with permission from Elsevier.} \label{fig_exp_img}
\end{figure}

Extensive studies over years have revealed three major cancer dormancy mechanisms: cellular dormancy, angiogenic dormancy and immunosurveillance \cite{Almog2010139,aguirre2007models}. On the cellular level, a tumor cell could be arrested at a certain stage of the cell cycle and unable to complete the cell division process successfully, resulting in a dormant solitary cell \cite{naumov2003ineffectiveness,aguirre2004green,kusumbe2009cancer}. On the cell population level, when the population does not gain enough ability to recruit blood vessels and promote neovascularization, the tumor cannot obtain sufficient nutrients necessary for its proliferation and as a result, angiogenic dormancy occurs \cite{hanahan1996patterns,Baeriswyl2009329}. On other hand, immunosurveillance operates when the immune system suppresses the proliferation of tumor cell population and leads to the dormancy of the tumor
\cite{krahenbuhl1976effects,weinhold1979tumor,matsuzawa1991survival,zou2005immunosuppressive,finn2006human}. Figure \ref{fig_exp_img}(a) shows an image of tumor tissue surrounded by immune cells. Figure \ref{fig_exp_img}(b) compares the morphology and vascular structure of dormant and fast-growing tumors.

A comprehensive understanding of cancer dormancy and the ``switch'' from a dormant to a proliferative state still needs to be strengthened. This is mainly due to the fact that efficient and accurate experimental or clinical approaches to track the states of individual cells in a dormant tumor {\it in vivo} throughout the entire dormancy period are still under development \cite{uhr1997cancer,marches2006cancer,quesnel2008dormant}.

Given the current need for further understanding of dormancy, computational modeling provides a powerful means to probe various scenarios for the underlying mechanisms. Specifically, modeling enables one to probe a variety of different dormancy scenarios by examining different combinations of mechanisms in order to see which ones provide possible explanations for experimental and clinical observations. Over the past few decades, computational modeling has played an important role in the study of the progression of solid tumors \cite{byrne2010dissecting}; a variety of models based on different mathematical schemes have been developed, including continuum models \cite{burton1966rate,greenspan1972models,owen2004mathematical,smallbone2005role}, discrete cell models \cite{jiang2005multiscale,quaranta2008invasion} and hybrid models \cite{anderson2005hybrid,kim2007hybrid}. Various models have been used to investigate cancer dormancy caused by cancer-immune interactions and other mechanisms, including ordinary differential equation-based models \cite{page2005mathematical,kuznetsov2001modeling}, stochastic differential equation-based models \cite{lefever1979bistability}, models based on kinetic theory for active particles \cite{de2008kinetic,brazzoli2010mathematical,bellomo2008mathematical}, and cellular automaton models \cite{takayanagi2006cellular}. However, the aforementioned studies neither explicitly demonstrated how the dynamic process of active proliferation after a certain period of dormancy emerges from various microscopic mechanisms nor showed the associated growth dynamics of the ``switch'' phenomenon. Therefore, predictive computational models that incorporate cellular-level microscopic mechanisms are needed to address these important issues.

In this paper, we generalize a two-dimensional (2D) cellular automaton (CA) model that we have devised to study proliferative growth of avascular solid tumors \cite{kansal2000cellular,schmitz2002cellular,torquato2011toward,jiao2011emergent,jiao2012diversity} in order to investigate tumor dormancy. Our goal is to formulate a dynamical model in which the ``switch'' to a proliferative state spontaneously emerges by incorporating additional interactions between the tumor and the microenvironment, for example the effects of immune system, which were not included in our previous CA model. The new rules of our CA model induce a ``competition'' between the tumor's propensity to proliferate and the microenvironmental factors that suppress its growth. In our model, a fraction of the dormant cells undergo phenotypic transformations triggered by intracellular factors or external stimulus and acquire the ability to actively proliferate. Subsequently, those microenvironmental factors act to suppress the growth of these transformed tumor cells either by killing some of these cells or turning these actively dividing proliferative cells back into dormant cells.

The ``competition'' between the tumor and the microenvironmental suppression factors either results in a ``stalemate'' for a period of time in which the tumor either eventually wins (spontaneously emerges) or is eradicated; or it leads to a situation in which the tumor is eradicated before such a ``stalemate'' could ever develop. Since we are mainly interested in the situations in which tumor growth involves a period of dormancy, we will henceforth focus on those situations in which a ``stalemate'' between the tumor and the microenvironmental suppression factors develops. Our model demonstrates that a variety of parameters characterizing the tumor-host interactions may greatly alter the growth dynamics of the tumor. These parameters include the rate of phenotypic transformation, by which the tumor cells gain the ability to proliferate against those suppression factors, the suppression rate imposed by suppression factors on individual tumor cells, and the mechanical rigidity of the microenvironment. The growth dynamics influenced by these parameters include the existence of a dormant period in tumor's growth, the length of the dormant period (if there exists one) and the existence of a sudden ``switch'' to a highly proliferative state. We also demonstrate that if the number of actively dividing cells within the proliferative rim reaches a critical, yet low level, the tumor has a high probability to begin rapid proliferation. While we study a 2D CA model for simplicity in this paper, our model can be easily generalized to three dimensions (3D).

\section*{Materials and Methods}
We divide the two-dimensional square simulation box into different polygonal units (i.e., automaton cells). Our model is coarse-grained, allowing us to grow the tumor from a very small size with a cross section of roughly 1000 real cells through to a fully developed tumor with a cross section consisting of $2.0 \times 10^6$ cells. Specifically, the innermost automaton cells represent roughly 100 real tumor cells or or a region of host microenvironment of similar size, while the outermost automaton cells represent roughly $10^4$ real tumor cells or a region of host microenvironment of similar size. To generate the automaton cells in the simulation box, we first fill the simulation box with non-overlapping circular disks (or spheres in 3D) using random-sequential-addition packing method \cite{torquato2001random} until there is no void space left for additional circular disks (or spheres in 3D). Periodic boundary conditions are used for generating the packing. Then we divide the simulation box into polygons (or polyhedra in 3D), each polygon (or polyhedron in 3D) associated with a particle center, such that any point within a polygon (or polyhedron in 3D) [i.e., a Voronoi polygon (or polyhedron in 3D)] is closer to its associated particle center than to any other particle centers. The resulting Voronoi polygons (or polyhedra in 3D) are referred to as automaton cells. In this paper, we focus on the two-dimensional case, but our model should be readily generalized to three dimensions.

The microenvironment surrounding a tumor is mainly composed of stroma cells and extracellular matrix (ECM). In the current model, we explicitly take into account the effects of the ECM macromolecule density, ECM degradation by the proliferative cells, and the pressure built up due to the ECM deformation by tumor growth. The effects of the stroma cells are not explicitly considered in our current model. Henceforth, we will refer to the regions of microenvironment as ECM-associated cells for simplicity. In addition, we consider the interactions between the tumor and the various suppression factors in the microenvironment, for example the immune system. Since we consider development of primary tumor or local recurrences of micrometastases under microenvironmental suppression, invasive tumor growth is not a mechanism relevant for our purposes and hence is not included in our dormancy model.

In our model of noninvasive proliferative tumor growth, tumor cells can be in one of the three possible states: proliferative, quiescent or necrotic, depending on their nutrient supply. Proliferative cells are tumor cells that have enough nutrients and possess the ability to divide. Quiescent (or arrested) cells are tumor cells that are alive, but do not have enough nutrient supply to support cell division. Quiescent cells can eventually become inert, necrotic (dead) cells due to an insufficient nutrient supply. In this paper, we focus on avascular tumor growth and assume that there is no explicit angiogenesis during the growth process (although this assumption can be relaxed). The nutrients available to tumor cells are those that diffuse into the tumor region through tumor edge. As the tumor grows, the amount of nutrient supply, which is proportional to the perimeter of the tumor interface (or surface area of the tumor interface in 3D), cannot meet the needs of all of the tumor cells. As a result, quiescent and necrotic regions emerge near the center of the tumor. The state of a tumor cell is determined by its distance to the tumor edge (i.e., the source of nutrients). We assume that proliferative cells more than $\delta_{p}$ away from the tumor edge become quiescent and quiescent cells more than $\delta_{n}$ away from the tumor edge become necrotic (see details below).

In this section, we will introduce our CA dormancy model, which modifies our previous basic CA models of tumor growth \cite{jiao2011emergent,jiao2012diversity,jiao2013evolution} by introducing several additional parameters to incorporate the interactions between tumor cells and the microenvironmental suppression factors. This dynamical model is capable of producing situations in which a ``switch'' from a dormant state to a proliferative state spontaneously emerges.

\subsection*{Noninvasive proliferative tumor growth}
We now specify the cellular automaton rules used in our model for noninvasive proliferative tumor growth. Each ECM-associated automaton cell is assigned a specific density $\rho_{\mbox{\tiny{ECM}}}$, representing the density of the ECM molecules within the automaton cell. If a proliferative cell divides, its daughter cell occupies a nearby ECM-associated cell. The daughter cell pushes away or degrades the ECM within the ECM-associated cell it occupies. Initially, a tumor is introduced by designating several automaton cells at the center of the growth permitting region as proliferative tumor cells. Then time is discretized into units, with each time step representing one day. At each time step, the tumor grows according to the following cellular automaton rules.
\begin{itemize}
\item Quiescent cells more than a certain distance $\delta_n$ from the tumor's edge are turned necrotic. The tumor's edge, which is assumed to be the source of nutrients, consists of all ECM-associated cells that border the tumor. The critical distance $\delta_n$ for quiescent cells to turn necrotic is computed as follows:
    \begin{equation}
    \label{eq_deltan} \delta_n = aL_t^{(d-1)/d},
    \end{equation}
where $a$ is the necrotic thickness controlled by nutritional needs, $d$ is the Euclidean spatial dimension and $L_t$ is the distance between the geometric centroid ${\bf x}_c$ of the tumor (i.e., ${\bf x}_c = \sum_i^N {\bf x}_i/N$, where $N$ is the total number of cells in the tumor) and the tumor edge cell that is closest to the quiescent cell under consideration.
\item Proliferative cells more than a certain distance $\delta_p$ from the tumor's edge are turned quiescent. The critical distance $\delta_p $ is given by
    \begin{equation}
    \label{eq_deltap} \delta_p = bL_t^{(d-1)/d},
    \end{equation}
where $b$ is the proliferative thickness controlled by nutritional needs, $d$ is the spatial dimension and $L_t$ is the distance between the geometric tumor centroid ${\bf x}_c$ and the tumor edge cell that is closest to the proliferative cell under consideration.
\item The probability of division for a proliferative cell used in our model is
    \begin{equation}
    \label{eq_complicated} p_{div} = {p_0}[1-\rho_{\mbox{\tiny{ECM}}}-\omega^*(\xi-1) + \xi\frac{\ell}{w}].
    \end{equation}
where $p_{0} = 0.192$ is the base probability of division linked to cell-doubling time, $\rho_{\mbox{\tiny{ECM}}}$ is the local ECM density, $\omega^*=2\rho_{\mbox{\tiny{ECM}}}^0$ is a parameter taking into account the effect of pressure, $\xi=\rho_{\mbox{\tiny{ECM}}}/\rho_{\mbox{\tiny{ECM}}}^0$ is the ratio of current average ECM density over the initial density, and $\ell$ and $w$ are, respectively, the length and width of local protrusion tips.
\end{itemize}

\subsection*{Interactions between the tumor and the microenvironmental suppression factors}
Here, we specify the additional interaction rules between the tumor and the microenvironmental suppression factors beyond the aforementioned ones for noninvasive proliferative growth, which were not included in our previous CA models. We assume that there are two possible states of proliferative cells, dormant or actively dividing, depending on their interactions with the microenvironmental suppression factors.
\begin{itemize}
\item Initially, we assume that all proliferative cells are kept in dormant states by the microenvironmental suppression factors, which means that they are not able to divide.
\item At each day, beyond the aforementioned CA rules for proliferative noninvasive growth, each dormant proliferative cell has a certain probability $\gamma$ to change in their phenotypes due to intracellular factors or external stimulus. The cell with phenotype change gains different degrees of resistance to the suppression factors in the microenvironment, depending on the specific phenotype change the cell undergoes. For example, mutated leukaemic cells in acute myeloid leukaemia acquire resistance to cytotoxic T lymphocytes-mediated cell lysis, whose degree is related to the level of the cell's expression of B7-H1 or B7.1 \cite{nc1}. For simplicity, we divide the phenotypic changes into two different types: \textit{weak} changes and \textit{strong} changes with respect to their resistance to the suppression factors in the microenvironment (i.e. their ability to actively proliferate). Henceforth, we will refer to these phenotypic changes as ``transformations'' and the cells that undergo these changes as ``transformed'' cells for simplicity. Strong-type ``transformed'' cells gain a larger competition advantage and thus have a greater ability to divide actively. The quantities $x_{W}$ and $x_{S}$ are the fractions of weak-type ``transformations'' and strong-type ``transformations''. Henceforth, we set $x_{W}$ 0.99 and $x_{S}$ 0.01.
\item At each subsequent day, the microenvironmental suppression factors will counteract the weak-type ``transformed'' and strong-type ``transformed'' cells with probabilities $\alpha_{W}$ and $\alpha_{S}$. The suppression factors in the microenvironment will either kill the ``transformed'' cells or turn them back into dormant cells \cite{krahenbuhl1976effects,weinhold1979tumor}.
\item When the number of tumor cells reaches a certain threshold $N_{T}$, strong reactions of the microenvironmental factors are triggered and those factors start to kill the ``transformed'' cells. The parameter $N_{T}$ is introduced to ensure that the tumor is not completely removed by the microenvironmental suppression factors. Note that the particular choice of $N_{T}$ barely has any effect on the simulation results within a relatively wide range of $N_{T}$ values.
In this work $N_{T}$ is set to be 50, a sufficiently small value that leads to biophysically  realistic outcomes.
As the tumor grows, the microenvironmental factors are weakened by the tumor, resulting in weaker suppression of the tumor cells \cite{bos2012treg,blatner2012expression}. Therefore, when the microenvironmental factors counteract the ``transformed'' cells, the fraction of the cells that are killed can be coupled with the growth rate of the tumor by
    \begin{equation}
    \label{eq_killing} k = k_{0}(1-\frac{1}{\triangle r_{C}}\frac{dA}{dt}).
    \end{equation}
where $k_{0}$ is a constant characterizing the strength of the suppression factors in the microenvironment, $dA/dt$ is the daily area change of the tumor (i.e. the growth rate of the tumor), and $\triangle r_{C}$ is the critical value of the tumor's growth rate. In this work, $\triangle r_{C}$ is chosen as half of the tumor's maximum growth rate under suppression, but our numerical tests have revealed that the simulation results are insensitive to the choice of $\triangle r_{C}$ as long as $\triangle r_{C}$ is smaller than the tumor's maximum growth rate. When the growth rate of the tumor reaches this critical value, the suppression factors become too weak to kill any actively dividing tumor cells and $k$ is set to be 0 \cite{bos2012treg,blatner2012expression}.
\item Due to the ``competition'' between the tumor and the suppression factors in the microenvironment, the ratio of the number of actively dividing proliferative cells over the total number of proliferative cells $n^{pro}_{acti}/n^{pro}$ changes with time. The larger is this ratio $n^{pro}_{acti}/n^{pro}$, the larger is the amount of nutrients the tumor tissue consumes. As a result, the nutrient concentration around the tumor depends on the ratio $n^{pro}_{acti}/n^{pro}$. Therefore, we make the necrotic thickness $a$ and proliferative thickness $b$ functions of $n^{pro}_{acti}/n^{pro}$:
    \begin{equation}
    \label{eq_nec_thickness} a = a_{0}[q - (q - 1.0)\frac{n^{pro}_{acti}}{n^{pro}}].
    \end{equation}
    \begin{equation}
    \label{eq_pro_thickness} b = b_{0}[s - (s - 1.0)\frac{n^{pro}_{acti}}{n^{pro}}].
    \end{equation}
where $a_{0} = 0.58$ mm$^{1/2}$ and $b_{0} = 0.30$ mm$^{1/2}$ are base necrotic thickness and base proliferative thickness respectively, $q=1.6,s=2.0$ are parameters determining the ranges of necrotic thickness and proliferative thickness as $n^{pro}_{acti}/n^{pro}$ changes.
\end{itemize}

\begin{table}[!ht]
\caption{\bf{Parameters characterizing the interactions between tumor suppression factors and tumor cells in the CA dormancy model. Note that the two ``critical threshold'' parameters themselves do not incorporate any additional CA rules.}}
\begin{center}
\begin{tabular}{>{\centering\arraybackslash}m{1cm}m{6cm}} \\ \hline\hline
\multicolumn{2}{c}{\textbf{Tumor growth parameters}} \\
\hline
$\gamma$ & Probability of phenotypic change for a dormant proliferative cell to acquire the dividing ability \\
\hline
$x_{W}$ & Fraction of weak-type transformations \\
\hline
$x_{S}$ & Fraction of strong-type transformations \\
\hline\hline \\
\multicolumn{2}{c}{\textbf{Microenvironmental suppression parameters}} \\
\hline
$\alpha_{W}$ & Probability that suppression factors counteract the weak-type transformed cell at each day \\
\hline
$\alpha_{S}$ & Probability that suppression factors counteract the strong-type transformed cell at each day \\
\hline
$k$ & Fraction of ``transformed'' cells killed when suppression factors counteract the ``transformed'' cells (time dependent)\\
\hline\hline \\
\multicolumn{2}{c}{\textbf{Critical threshold values}} \\
\hline
$N_{T}$ & Critical value of proliferative tumor cell number, beyond which suppression of tumor growth is triggered \\
\hline
$\triangle r_{C}$ & Critical value of tumor growth rate, beyond which the suppression factors are unable to kill the ``transformed'' cells \\
\hline\hline
\end{tabular}
\end{center}
\label{tab_Param}
\end{table}

The aforementioned additional parameters associated with the new rules that we employ for dormancy (beyond the ones for noninvasive proliferative growth) are summarized in Table \ref{tab_Param}. These parameters are sufficient to formulate a model in which the transition from ``dormant'' to proliferative state emerges spontaneously. Note that unlike other parameters listed in Table \ref{tab_Param}, the two critical threshold parameters themselves do not incorporate any additional CA rules. Instead, the critical threshold parameters determine when the microenvironmental suppression factors are able to kill the proliferative cells. Also, it is noteworthy that we map the complicated tumor-host interactions onto a number of ``effective'' parameters. The values of these parameters could differ for different tumors in different microenvironments. It is noteworthy that currently due to a lack of detailed in-vivo or in-vitro data for the growth dynamics of a dormant tumor, we are not able to determine the values of the parameters in our model for a specific real system. Instead, we have done a full parametric study to probe different outcomes corresponding to different parameter values in the subsequent sections. However, once we obtain the statistics of a dormant tumor as a function of time from the initiation of the tumor, we should be able to extract the parameter values for the tumor by fitting the statistics. At this stage, the extracted parameter values could be applied to other tumors of similar type.

\subsection*{Noninvasive proliferative tumor growth under suppression}
Here, we specify how the additional interaction rules are coupled together with the original CA rules for noninvasive proliferative tumor growth, resulting in noninvasive proliferative tumor growth under suppression.
\begin{itemize}
\item As mentioned above, proliferative cells in the dormant state do not divide. Only proliferative cells in actively dividing states actually proliferate.
\item At each day, each dormant proliferate cell is checked to see if it enters the active state according to the interaction rules. Once it begins to actively divide, it proliferates according to the CA rules for proliferative tumor growth.
\item At each day, each active proliferative cell is checked to see if it is killed or turned back into dormant cell according to the interaction rules.
\item Quiescent cells and necrotic cells act according to CA rules for proliferative tumor growth. However, the values of parameters $a$ and $b$ determining the transitions from necrotic cells to quiescent cells and from proliferative cells to quiescent cells respectively are influenced by interaction rules, as mentioned above.
\end{itemize}

Note that our CA dormancy model should be readily generalized to angiogenic dormancy by explicitly considering the angiogenic process and vascular tumor growth. This is beyond the scope of this work and will be addressed in future work.

\section*{Results}
In this section, we apply our CA model and show that it produces a dormancy period of the tumor that can lead to a subsequent emergent ``switch'' behavior to a proliferative state. A homogeneous distribution of ECM density is used for simplicity \cite{jiao2011emergent}. A circular growth permitting region containing $\sim 2\times10^4$ automaton cells is employed. Simulating the growth of a 2D tumor from several cells (representing roughly 1000 real cells) to a macroscopic-size tumor (a cross section of 5 $cm^2$ consisting of $\sim 2\times10^6$ real cells) with a period of dormancy up to a person's life ($\sim 80$ years) generally takes no more than a few minutes on a standard Dell Workstation (Precision T3400).

Initially, a few automaton cells at the center of the growth-permitting region are designated as proliferative cells. Then the initial tumor is allowed to grow according to our CA model incorporating the additional interaction rules between the tumor and the suppression factors in the microenvironment. Certain geometrical characteristics of the tumor (e.g., tumor area, areas of different tumor cell populations) and its morphology (e.g., the geometrical positions of the tumor cells) are collected every $T_c$ days. We set $\gamma=0.005$, $\alpha_{W}=0.75$, $\alpha_{S}=0.15$, $k_{0}=0.8$ and use these parameter values throughout this paper, except where otherwise stated.

\subsection*{Statistics of tumor growth}
\begin{figure}[!ht]
\begin{center}
$\begin{array}{c}
\includegraphics[width=0.45\textwidth]{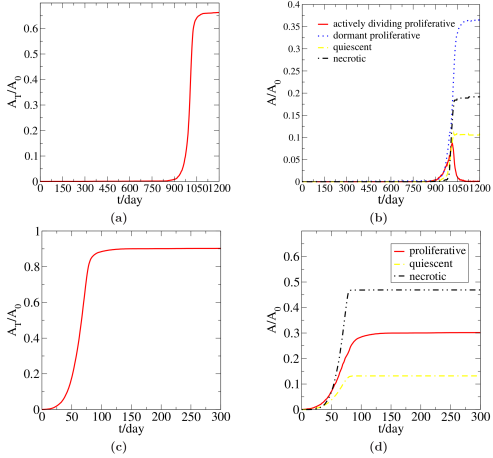}
\end{array}$
\end{center}
\caption{(Color online) Upper panel: statistics of a simulated noninvasive tumor growing in the ECM with ${\bar \rho}_{\mbox{\tiny ECM}}=0.3$ and microenvironmental suppression factors, as predicted by the ``CA dormancy model''. (a) Tumor area $A_{T}$ normalized by the area $A_{0}$ of the growth permitting region. (b) Areas of different cell populations normalized by the area $A_{0}$ of the growth permitting region. Lower panel: statistics of a simulated noninvasive tumor growing in the ECM with ${\bar \rho}_{\mbox{\tiny ECM}}=0.3$ without suppression. (c) Tumor area $A_{T}$ normalized by the area $A_{0}$ of the growth permitting region. (d) Areas of different cell populations normalized by the area $A_{0}$ of the growth permitting region. } \label{fig_model}
\end{figure}

Here we consider the growth of a proliferative tumor in a confined space with ${\bar \rho}_{\mbox{\tiny ECM}}=0.30$. As shown in Figure \ref{fig_model}(a), with the interactions between the tumor and the microenvironmental suppression factors incorporated, there exists a period of dormancy in the tumor's growth. Specifically, for the initial approximate 900 days, the tumor stays in a dormant state. Suddenly at approximately day 900, the tumor switches its behavior and begins rapid proliferation. The virtual patient would die 100 days after this critical point in time. Figure \ref{fig_model}(b) shows the areas $A$ of different populations normalized by the area of the growth-permitting area $A_{0}$. For purposes of comparison, Figure \ref{fig_model}(c) and Figure \ref{fig_model}(d) show the statistics of the tumor growth without the suppression of microenvironmental factors. Moreover, by comparing Figure \ref{fig_model}(a) and Figure \ref{fig_model}(c), it is seen that the interactions between the tumor and the microenvironmental suppression factors lead to the existence of a dormancy period and a subsequent emergent ``switch'' behavior of the tumor from a dormant state to a proliferative state. Also, from the comparison of Figure \ref{fig_model}(b) and Figure \ref{fig_model}(d), one can see that the additional interaction rules alter the fractions of necrotic cell population and proliferative cell population within the tumor. When suppression of the tumor growth is present, the necrotic region decreases and the proliferative region increases relatively; the area of the quiescent region remains almost unchanged.

\begin{figure}[!ht]
\begin{center}
$\begin{array}{c}
\includegraphics[width=0.45\textwidth]{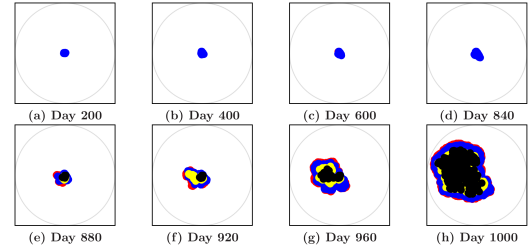}
\end{array}$
\end{center}
\caption{(Color online) Snapshots of a simulated noninvasive tumor growing in the ECM with ${\bar \rho}_{\mbox{\tiny ECM}}=0.3$ on different days given by the CA dormancy model. Upper panel: Dormancy period. Lower panel: Regrowth period.} \label{fig_snapshot}
\end{figure}

Figure \ref{fig_snapshot} shows snapshots of the simulated 2D tumor. It can be clearly seen that the tumor develops a highly aspherical morphology due to the interactions between the tumor and the microenvironmental factors. Figure \ref{fig_snapshot} also demonstrates that the tumor hardly grows during the period of dormancy, but once the ``switch'' occurs, the tumor expands very rapidly. Henceforth, we will use the ``CA dormancy model'' to investigate the effects of the various parameters characterizing the tumor-host interactions on the growth dynamics of the tumor. These parameters include the rate of phenotypic transformation, by which the tumor cells gain the ability to proliferate against those suppression factors, the suppression rate imposed by suppression factors on individual tumor cells, and the mechanical rigidity of the microenvironment.

\subsection*{Suppression rate vs transformation rate}
\begin{figure}[!ht]
\begin{center}
$\begin{array}{c}
\includegraphics[width=0.45\textwidth]{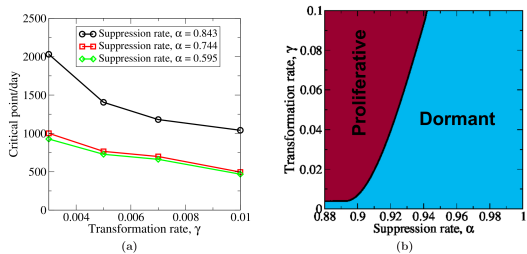}
\end{array}$
\end{center}
\caption{(Color online) (a) The ``critical'' point at which the noninvasive tumor growing in the ECM with ${\bar \rho}_{\mbox{\tiny ECM}}=0.3$ switches from a dormant state to a proliferative state as functions of $\alpha$ and $\gamma$. (b) A schematic phase diagram that characterizes the growth dynamics of a noninvasive tumor growing in the ECM with ${\bar \rho}_{\mbox{\tiny ECM}}=0.3$ under different $\alpha$ and $\gamma$.} \label{fig_phase_diagram}
\end{figure}

Here we investigate growth dynamics of the tumor under different suppression rates $\alpha$ and phenotypic transformation rates $\gamma$. The suppression rate $\alpha$ is defined as the following weighted average:
\begin{equation}
    \alpha = \alpha_{W}\cdot x_{W}+\alpha_{S}\cdot x_{S}.
\end{equation}
where $x_{W}=0.99$ and $x_{S}=0.01$ are the fractions of weak-type ``transformations'' and strong-type ``transformations'', and $\alpha_{W}$ and $\alpha_{S}$ are the suppression rates of the weak-type ``transformed'' cells and strong-type ``transformed'' cells by the microenvironmental factors.
It is found that increasing $\alpha$ and decreasing $\gamma$ generally increases the length of the dormancy period and delays the ``switch'' from a dormant state to a proliferative state, as demonstrated in Figure \ref{fig_phase_diagram}(a). Within some regimes of $\alpha$ and $\gamma$, the tumor could lie dormant for a period equal to or longer than a person's life ($\sim 80$ years).

Based on our simulation results, we construct a ``phase diagram'' to characterize the tumor's growth dynamics in terms of $\alpha$ and $\gamma$, as shown in Figure \ref{fig_phase_diagram}(b). There are two regions in this phase diagram: proliferative and dormant regions. By ``proliferative'', we mean that the tumor resumes rapid proliferation after a period of dormancy and the length of the dormancy period is less than a virtual patient's life; by ``dormant'', we mean that the tumor remains in a dormant state during the whole life of a person and undetected by conventional clinical methods (usually clinicians call such tumors ``benign'' \cite{hoffman1980clinically,mahaley1989national,england1989localized}). The solid line separates the two regions, and crossing this boundary line is associated with a ``phase transition''.

\subsection*{Rigidity of the microenvironment}
\begin{figure}[!ht]
\begin{center}
$\begin{array}{c}
\includegraphics[width=0.45\textwidth]{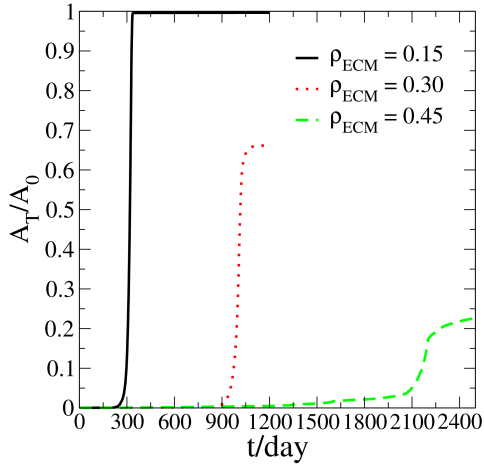}
\end{array}$
\end{center}
\caption{(Color online) Simulated tumor area $A_{T}$ normalized by the area $A_{0}$ of the growth permitting region of a noninvasive tumor growing in the ECM with different ${\bar \rho}_{\mbox{\tiny ECM}}$.} \label{fig_rig}
\end{figure}

Various mechanical cues in the microenvironment could influence the growth dynamics of the tumor \cite{nc2}. Here we only consider the effects of the ECM macromolecule density, ECM degradation by the proliferative cells, and the pressure built up due to the ECM deformation by tumor growth. As shown in Figure \ref{fig_rig}, when ECM rigidity increases, the time at which the switch occurs gets delayed significantly and the final size of the tumor when it plateaus appreciably decreases. For example, with all the other parameters fixed, when the tumor grows in a soft ECM with ${\bar \rho}_{\mbox{\tiny ECM}}=0.15$, the ``switch'' point occurs approximately on day 250. This is to be contrasted with growth in a rigid ECM with ${\bar \rho}_{\mbox{\tiny ECM}}=0.45$ where the dormancy period could last for 2,000 days before a ``switch'' to a proliferative state occurs. Also, the plateau size of the tumor growing in a soft ECM with ${\bar \rho}_{\mbox{\tiny ECM}}=0.15$ is five times as large as that of one growing in a rigid ECM with ${\bar \rho}_{\mbox{\tiny ECM}}=0.45$. In other words, when the tumor grows in a harsher microenvironment, it's harder for the tumor to break out of a dormant state and potential proliferative growth is largely suppressed. Note that the rigidity of the microenvironment could also affect tumor growth via various intracellular signaling processes \cite{nc2}. Those mechanotransduction effects will be incorporated into our CA dormancy model in future work, which could result in different scenarios from those reported here \cite{nc2}.

\subsection*{Strength of the suppression factors}
\begin{figure}[!ht]
\begin{center}
$\begin{array}{c}
\includegraphics[width=0.45\textwidth]{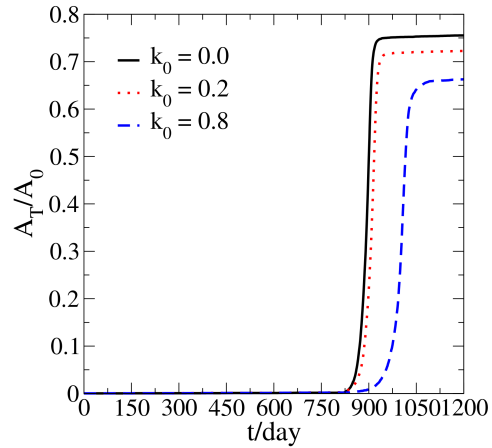}
\end{array}$
\end{center}
\caption{(Color online) Tumor area $A_{T}$ normalized by the area $A_{0}$ of the growth permitting region of a simulated noninvasive tumor growing in the ECM under different killing rates by microenvironmental suppression factors. The parameter $k_{0}$ is the fraction that the suppression factors from the microenvironment kill the actively dividing proliferative cells when the suppression factors counteract these cells.} \label{fig_killling_rate}
\end{figure}

Here we investigate how tumor growth dynamics changes with the strength of the microenvironmental suppression factors. As shown in Figure \ref{fig_killling_rate}, increasing the fraction of actively dividing tumor cells that are killed [i.e., increasing $k_{0}$ in the equation (\ref{eq_killing})] when the microenvironmental suppression factors (which we recall could either kill the ``transformed'' cells or turn them back into dormant cells) delays the ``switch'' point from a dormant state to a rapid proliferative state and decreases the final tumor size. However, relatively speaking, the simulated tumor growth statistics are insensitive to $k_{0}$ compared to the influences of the aforementioned other factors. Note that even when the suppression factors can only turn the ``transformed'' cells back into dormant cells and do not kill any ``transformed'' cells (i.e. $k_{0}=0$), a ``switch'' behavior from a dormant state to a rapid proliferative state can still emerge. This indicates that turning the active proliferative cells back into dormant cells could also be a possible independent mechanism leading to a dormancy period and a subsequent ``switch'' to a proliferative state.

\section*{Discussion}
In this paper, we generalized a two-dimensional cellular automaton (CA) model previously developed for proliferative growth of avascular solid tumors to investigate tumor dormancy and evasion from dormancy to proliferation. Our CA dormancy model incorporates a variety of cell-level tumor-host interactions, including those between the tumor and the suppression factors in the microenvironment, for example the immune system. Our CA dormancy model induces a ``competition'' between the tumor's propensity to proliferate and the microenvironmental factors that suppress its growth. Our CA dormancy model predicts a dramatic emergent ``switch'' behavior from a dormant state to a rapidly proliferative state. Our results show that under the suppression of microenvironmental factors, the tumor develops a highly aspherical morphology with an larger proliferative region and a smaller necrotic region than those of a tumor that grows without the presence of suppression factors. We also predict that if the number of actively dividing cells within the proliferative rim of tumor reaches a critical, yet low level, the tumor has a large probability to resume rapid regrowth and exit dormancy. In addition, we demonstrate that a variety of different factors could greatly alter tumor growth dynamics, including the rate of phenotypic transformations, the suppression rate by the microenvironmental factors, the mechanical rigidity of the microenvironment, and the strength of the suppression factors. However, relatively speaking, the tumor growth is insensitive to the strength of the suppression factors in terms of killing active proliferative cells. We inferred from our simulation results a qualitative phase diagram to characterize the growth dynamics of the tumor under the suppression of microenvironmental factors in terms of the phenotypic transformation rate and the suppression rate. In this paper we focused on the two-dimensional case, but our model should be easily generalized to three dimensions.

At the cellular level, the origin of the ``stalemate'' between the tumor and microenvironmental suppression factors remains unclear. This ``stalemate'' may come from cell proliferation balanced by cell death, which could be the case for a dividing cancer stem cell \cite{nc3}. Arrested tumor cell proliferation imposed by microenvironmental factors could also result in the ``stalemate'' between the tumor and the microenvironmental suppression factors. Both scenarios could account for the case of differentiated cancer cells, since our CA dormancy model is coarse-grained and therefore considers the effective behavior of the tumor.

Our CA dormancy model may shed light on the fundamental understanding of cancer dormancy phenomenon. Specifically, our CA dormancy model proposes possible scenarios for cancer dormancy that during the dormancy period the great majority of proliferative cells stay in a dormant state, while only a small portion of proliferative cells, i.e., ``transformed'' cells are actively dividing, and the microenvironmental suppression factors counteract these ``transformed'' cells by either killing them or turning them back into dormant cells. As a result, the tumor cell population is barely expanding during the dormancy period. It is noteworthy that our CA dormancy model predicts that the tumor either eventually spontaneously emerges or is eradicated after a period of a ``stalemate'' between the tumor and the microenvironmental suppression factors; or the tumor is eradicated before such a ``stalemate'' could ever develop. These predicted scenarios arising from the interaction between the tumor and the microenvironmental suppression factors in our simulation qualitatively match the experimental observations of the cancer immunoediting process, by which the immune system controls the tumor growth and necessarily leads to tumor escape or elimination \cite{Matsushita2012}. The predictions of our CA dormancy model can be further verified by comparing the macroscopic geometrical and dynamical properties of our simulated tumor in different microenvironments \cite{kloxin1} to those obtained by experimental data from future animal studies. In future work we plan on incorporating recently discovered mechanisms for cancer dormancy via the clinical trials and experiments \cite{nc4, nc5} to better inform our computational model. These results together could aid in answering the important fundamental question of whether the majority of cancer cells in a dormant tumor are arrested at a certain stage of the cell cycle or not. Furthermore, they will have significant treatment implications in terms of what stage of the cell cycle the therapies should target \cite{wells2013}.

Besides the aforementioned influences, our findings informed by clinical data might be able to provide further insights to novel early cancer detection and therapy. For example, a new cancer drug that suppresses the emission of CD47 by the tumor tissues, which helps the tumor cells evade attack by the immune system, has been discovered \cite{willingham2012cd47}. It was shown via {\it in vitro} experiments that this drug is able to kill a variety of cancer cell types. Thus, an effective clinical application of this drug depends upon the ability to identify different tumor cell populations while they are dormant. Our work may serve to provide insights to the application of this new drug as well by contributing to the development of new early detection methods. In addition, our work may shed light on why the immune system may not always be able to prevent tumor progression. Specifically, our work shows that even if the immune system maintains its strength throughout the tumor growth process, there is still a high possibility that the immune system could eventually fail, which is to be contrasted with the simple explanations that it becomes weaker as the tumor develops \cite{weinhold1979tumor,matsuzawa1991survival,zou2005immunosuppressive,finn2006human}. Also, for a tumor of a specific type, we can extract the parameter values in our model by fitting our simulation results to the statistics of a real in-vitro or in-vivo tumor of this type. Then we can utilize our model to explore optimal treatment strategies for the tumors of this specific type. In addition, once we determine the effects of a specific microenvironmental factor (e.g., specific integrins \cite{nc6}) on the parameter values in our model, we could then study the effects of this microenvironmental factor on the tumor growth dynamics.

Our current CA dormancy model is still preliminary, and to achieve our ultimate goal of understanding cancer dormancy and progression, we need to develop robust models that incorporate appropriate cell-level tumor-host interactions that are informed by experiments. For example, by explicitly considering angiogenesis and using more realistic distribution of ``transformed'' tumor cells' resistance to microenvironmental suppression factors (currently we just divide the ``transformed'' cells into two types with respect to their resistance to microenvironmental suppression factors: \textit{weak} and \textit{strong}), our model might be able to yield more realistic results and improve our understanding of cancer dormancy and progression. Also, the effects of tumor cell competition, cooperation and the microenvironmental changes caused by tumor cell activities could be incorporated to further strengthen our CA dormancy model \cite{nc7}. It is noteworthy that although we employ interaction rules based on a discrete cell model to describe ``competition'' between the tumor and the microenvironmental suppression factors, alternatives such as evolutionary game theory implemented by partial-differential equations are also available to address the interplay between the tumor and the microenvironment \cite{nc8}.

\section*{Acknowledgments}
S. T. is grateful to Paul Davies for bringing his attention to the cancer dormancy problem. The research described was supported by the National Cancer Institute under Award NO. U54CA143803. The content is solely the responsibility of the authors and does not necessarily represent the official views of the National Cancer Institute or the National Institutes of Health.



\clearpage


\end{document}